# Success factors for Crowdfunding founders and funders

Yang Song; Robert van Boeschoten

**Abstract**

Crowdfunding has been used as one of the effective ways for entrepreneurs to raise funding especially in creative industries. Individuals as well as organizations are paying more attentions to the emergence of new crowdfunding platforms. In the Netherlands, the government is also trying to help artists access financial resources through crowdfunding platforms. This research aims at discovering the success factors for crowdfunding projects through crowdfunding platforms from both founders' and funders' perspective. We designed our own website for founders and funders to observe crowdfunding behaviors. Our research will contribute to crowdfunding success factors related to issues of trust and decision making and provide practical recommendations for practitioners and researchers.

**Introduction**

Entrepreneurs are using social media such as LinkedIn, Facebook and Twitter to get access to information and resources (Song & Vinig, 2012). They usually find it difficult to raise funding during the early stage of their entrepreneurial process (Cosh, Cumming, & Hughes, 2009). Most of them are not eligible for bank loans or debt due to the fact that their businesses' operating history is limited. Some entrepreneurs are able to get funding from "family, friends and fools" (from the so-called three Fs) but in many cases this is not efficient (Mason, 2007). The emergence of social media provides a different platform for entrepreneurs to raise funding. In particular, they start seeking for financial help from the general public (the "crowd") instead of approaching financial investors such as business angels, banks or venture capital funds(Schwienbacher,A., Larralde,B.., 2010).

The interaction between founders and funders interested in crowdfunding can be organized through crowdfunding platforms. The amount of platforms is growing every day. Institutional organizations such as the European Commission and a wide range of European states are proactively interested in regulating all these platforms. Public funding institutions are collaborating with crowdfunding platforms in order to support creative workers get access to financial resources for their projects. Crowdfunding seems like a promise to finance the future, especially that of the creative industries.

Crowdfunding, which typically involves collecting small amounts of money from a large number of people (Ordanini, Miceli, Pizzetti, & Parasuraman, 2011), is being used as an alternative way of financing projects, in particular for artists, cultural practitioners, designers, programmers, researchers, and small creative social entrepreneurs. All of them can be called as the potential founders for crowdfunding projects. Correspondingly, there are also funders for each project. However, the internal crowdfunding mechanisms such as how the funders make decisions for a project are beyond our knowledge. Unlike business angels or venture capital funds, crowd funders might not have any special knowledge about the industry nor the crowdfunding project (Belleflamme, Lambert, & Schwienbacher, 2014). This means that a relationship between funders and founders is solely based on the interaction facilitated by the platforms or other social media. The way the two parties can find agreement on what to contribute and find interest in each other as ground for developing a project, normally takes place in the online environment of the platform.

In this paper, we would like to discover the mechanisms that both founders and funders might come across during a crowdfunding process, we would like to provide practical recommendations for both founders and funders. First of all, we would like to identify opportunities and challenges from the founders' perspectives; second, we would like to help founders raise the interests of the funders for a project. Last but not least, we are interested in identifying the success factors of crowdfunding projects, in particular, we are interested in factors related to trust and transparency during a crowdfunding process. We aim at solving the issues during a crowdfunding process by answering the following questions: 1) How to find the right crowd for founders during crowdfunding process? 2) How to identify the interests of funders? How to test the involvement of the funders? 3) How to improve the trust between founders and funders through crowdfunding platforms? 4) During the entire crowdfunding process, why do a few projects succeed while other fail? 5) How can crowdfunding founders and funders build up efficient and strong community through crowdfunding platforms?

**Literature Review**

Crowdfunding is described as "an open call, essentially through the Internet, for the provision of financial resources either in form of donation or in exchange for some form of reward and/or voting rights in order to support initiatives for specific purposes" (Belleflamme et al., 2014; Schwienbacher,A., Larralde,B.., 2010). In other words, crowdfunding typically involves collecting small amounts of money from a large number of people, which is a new label for an activity that has a rich history in many domains (Ordanini et al., 2011). Crowdfunding success appears to be linked to products or services quality, in that products or services that signal a higher quality level are more likely to be funded especially in non-profit organizations or creative sectors (Schwienbacher,A., Larralde,B.., 2010). Crowdfunding platforms are a novel place for fundraising activities, functioning as online intermediaries between entrepreneurs with ideas and the public with money and expertise (Massolution, 2012).

Previous research shows that there are mainly four crowdfunding models: 1) the donation-based or patronage model, for philanthropic or sponsorship purposes; 2) the lending based model, as a peer-to-peer and peer-to business loans; 3) the reward-based model; for non-monetary rewards that are normally the result of the entrepreneurial activity; 4) the equity model, for financial and participation return (Mollick, 2014). Depending on the crowdfunding project, the founders need to choose a crowdfunding model. Besides crowdfunding models, there are also other factors that might be relevant to the success of crowdfunding project such as industries, social networks, trust etc. Therefore, we will discuss the relevant crowdfunding success factors in the following part.

*Creative industries*

The rise of the crowdfunding industry over the past decade comes from the advancement in web and mobile-based web applications and services (De Buysere,K., Gajda,O., Kleverlaan,R., Marom,D., 2012). In other words, we call it creative industries in our research. The term 'creative industries' includes a diverse range of sectors that are commonly thought of as being quite distinct from each other. Based on previous research, we identify several creative sectors, which have the highest possibility to raise funding through crowdfunding. The sectors include: advertising; architecture; the art and antiques market; crafts; design; designer fashion; film and video; interactive leisure software (such as computer games); music; the performing arts; publishing; software and computer services; and television and radio (Green, Ian Miles, & Rutter, 2007; Howkins, 2001).

*Social networks*

Networking is strongly related to entrepreneurship, which is "the process by which individuals - either on their own or inside organizations – pursue opportunities without regard to the resources they currently control" (Stevenson & Jarillo, 1990). Entrepreneurs might meet with an obstacle when they are about to use their resources. As a consequence, they have to take advantage of their existing social networks or try to establish new relationships. It has been long known that good social networks are one of the well-known advantages in business world especially in entrepreneurship (Schwienbacher,A., Larralde,B.., 2010).

The founders of crowdfunding projects, are a type of entrepreneurs who use social networks to communicate with their possible investors or funders. The terminologies of "crowdfunding" only emerged recently. However, the process of crowdsourcing through crowd or social networks has been used since 1714 when the British government offers a "Longitude Prize" of £20,000 for a reliable method of calculating a ship's longitude (Ross, 2012). With the development of social media and online social networks, crowdfunding became one of the ways for entrepreneurs to raise funding through their online social networks.

*Trust*

Within the context of social networks and collaborations, trust is an essential feature that has barely been addressed. Previous research shows that the dynamics of online cooperation has to do with social network size and trust (Meadows, 2007). One could open up to others because one feels more intimate during face-to-face communication in offline networks. The Internet has been a utopian place of regaining trust among human beings. Internet could enable self-governance and bring people with common interests together (Morozov, 2012; Rheingold, 1993). Although over the last decade many signs have been given that the Internet is not such an open space where everyone can create their commons but that there is actually a lot of control either by governments or by companies (Deibert, 2008; Goldsmith & Wu, 2006), the belief in an open Internet still remains (Keymolen, Prins, & Raab, 2013). The Internet is also an excellent space for entrepreneurship (Song & Vinig, 2012). The internet can offer entrepreneurs direct access to their customers without troublesome regulations. Therefore, a new economy can be built upon online communication, in which trust among strangers is one of the key values (Bottsman & Rogers, 2010).

Our research is to find out how this belief in online trust can be grounded on relationships that are experienced as trustworthy. Based on a study by Carolien Nevejan (2007) online interactions can be organized in the way we are made present in the interaction. Presence occurs in time, space and the acknowledgement of the persons in the interaction. All these categories demand in online networks to be designed. Interactions are often not direct online and the organization of space (as in the webpages) has implications to what is made relevant in the interaction. Webdesign does a lot of research into what attracts the visitor of a site on a page. Where does the attention go to and what does happen afterwards. This is the field of human - computer interaction design. Often the interaction is not just person to person but e.g. one to many or many to many which also requires a specific design to be successful. In Crowdfunding these design issues that are relevant for the creation of trust are done by the crowdfunding platforms. They have created a business canvas as a model for all kinds of fundraising activities.

*Crowdfunding Platform*

Crowdfunding platforms are a novel place for fundraising activities, functioning as online intermediaries between entrepreneurs with ideas and the public with money and expertise (Massolution, 2012). In other words, a crowdfunding platform is one of the interfaces between founders and funders. The platform seems to be a neutral interface as a business model, however, as a social model of building relationship between founders and funders, this is not the case. Both parties are troubled in making the platform a reliable interaction interface. We can judge the reliability of the communication through a face-to-face communication. However, through crowdfunding platform, this is completely different.

This behavior is not only done by founders and funders, the platform itself is active in building relationships through the interaction design choices it offers. If we look at what kind of actions are undertaken by the platforms, we see a variety of options regulated by their business canvas as blueprint for the interaction online. The capacity to act and receive feedback on your actions is fundamental for feeling present in the interaction. Trustworthy communication in online networks is based on configurations that are related to these dimensions. Some configurations create trust and others don't. The kind of interaction framed by the so called canvasses that are in use by the crowdfunding platforms create one kind of configuration. There seems to be a consensus on how things should be done although the nature of these relationships between founders and funders varies a lot. It all depends on what kind of agreement is made (action). Is this about equity then we presume a lot of trust needs to be guaranteed since the relationship is supposed to last a lot longer (time). When an agreement needs to be made on a gift for a social action then trust is less important since it is usually done in the spur of the moment. It is interesting to see how the decision to participate is being made.

Founders make a video to promote their product as a collaborate work of a group that addresses an unknown audience. They also show their contact address for personal information. Funders respond to the video by donating money or buying a product or clicking to a next proposal. The platform shows itself as a wall of proposals that seem similar but looked upon more closely vary enormously. Professional video's are interspersed with amateur material. Money is asked for donation or for founding a company by buying equity, all within the same webspace and with the same human-computer interface. Trust is being organized in the same way without regards for the difference in interaction needed to build a sustainable relationship between founders and funders.

The platforms are open for anyone to visit and participate while nowadays many voices call for a more secured space when it comes to financial transactions. This openness seems less likely when we hear about stories on surveillance techniques used by governments and business alike. Our profile is being soled and our opinions checked all the time. The Internet becomes a heavily filtered environment that leads to techno-regulations that are so far not part and parcel of crowdfunding platforms (Keymolen et al., 2013). This is all needed if we want to stimulate the possibilities of online commerce. The main problem with this issue is based on the fact that doing business is usually seen as a something that follows rules that are everywhere applicable. The interaction and process seems to be independent of the media/techniques used. That is also the point where it always turns out wrong and to which a lot of governmental regulations is addressed nowadays..

**Research Design**

Our research is conducted by three steps. In the first step, we interviewed the Dutch crowdfunding founders about their crowdfunding experiences. We used semi-structured interview and interviewed 8 crowdfunding founders. For each founder, we spent approximately one hour. Based on the findings from the interviews we made an online survey to further check the preferences of founders. There were 60 founders attending the survey.

In the second step, based on the results from the first step, we designed our tookit for crowdfunding founders. This toolkit can help the founders to analyze the best model for their crowdfunding project. We invited artists, cultural practitioners, designers, programmers, researchers, and small creative social entrepreneurs to use our toolkit. The toolkit is only useful for projects that aim to raise less than 50,000 euro. It is not intended to assists multi-million-euro startups and businesses and is directed to participants from the Netherlands. Besides, we also considered the amount of time and effort required to work on a crowdfunding project, as well as the reliance on friends and family for money etc.

In the third step, we designed a survey especially for crowdfunding funders. In this survey, we aim at discovering the motivations for funders to join a crowdfunding project. In particular, we are interested in how to build up trust between funders and founders during crowdfunding process. Therefore, we can better inform potential founders about what the crowdfunding process involve before they start their own projects. We sent out our survey link to the Dutch crowdfunding funders through emails and social media.

**Data description**

We conducted 8 in depth interview on Dutch crowdfunding founders. For each founder, we spent approximately one hour. Based on the findings from the interviews we made an online survey to further check the preferences of founders. There were 60 founders attending the survey for founders. We concluded our preliminary results about founders based on these 60 participants. Unfortunately, due to the time limitation, we only managed to invite 14 funders attending our survey for funders. Since previous research most focus on the perspectives of founders, we would like to highlight our findings about crowdfunding funders.

According to our survey in the appendix, we have 17 questions for crowdfunding funders. We found pre-purchase model accounts for 46% of all the funders. Compared to online communications in terms of social media, email  etc, about 54% of the funders prefer meeting with the initiators of the projects in person. Even though they would like to know them in person, they still want to interact with the founders through emails or crowdfunding platform.

We are still unsure about the motivation of the funders in order to make their decisions for a crowdfunding projects. However, 85% of the funders are interested in the story behind the crowdfunding project. 54% funders agree that the participation in the crowdfunding project is important. The personal interests and potential of the projects are two important factors for funders to select a crowdfunding project. To our surprise, the funders seldom contact the founders of the projects. Even worse, they never contact the other funders. We presume that the frequent social activities might help them improve the success rate of a crowdfunding project.

**Preliminary results**

This paper is based on a pilot study about crowdfunding founders and funders. We find mainly founders from creative industries are more likely to start up a crowdfunding project. The following results are concluded based on our different approaches.

*Insights from crowdfunding founders*

First，according to the insight of the founders, crowdfunding is more than just raising money. During the crowdfunding process, the founders can build their own community, which can be used to validates ideas or test, change and improve the projects. Meanwhile, the funders can also help the founders to spread the ideas through their networks. Crowdfunding can be very effective to organize one's personal network such as relatives and friends in other countries. Since fundraising is one of the main issue for startups, crowdfunding can make the founders feel more independent compared to traditional forms of finance such as VC funds and banks. However, they had to rely on their personal network during the whole crowdfunding process. Both the founders and the funders need to make great efforts to break the circle in order to attract more funders. As for time, the founders unilaterally agreed that they had to spend around two days a week solely on managing the campaign, spanning from one to three months. This proves to be particularly difficult to manage for those individual founders. Furthermore, promoting their project in front of crowd can not only often raise the expectations of their project but also put a lot of pressure on success.

During the interview, we also found most of the founders are lack of expertise in how to crowdfund, how to campaign and reach a wider audience. Opting for one crowdfunding platform instead of another was usually the result of either the platform's popularity or the referrals from colleagues. Our preliminary research also shows that most of the founders are unlikely to use crowdfunding platforms next time. This is because of 1) the big investment of time and effort and 2) the uncomfortable feeling of 'begging for money' from close family and friends. 3) Crowdfunding is not sustainable – most would not do it again unless they are sure they had something of concrete value to offer. The relationship with their audiences becomes more fragile when they are asking for money before offering the outcome.

In order to further understand the dynamics of crowdfunding, we used a survey involved in 12 questions. We found most of the founders invested 10 and 15 hours a week. Only 10% of all respondents spent more than 30 hours a week. Most of the time was spent on promoting the project. We concluded from the founders that prompting the project is one of the most demanding part for crowdfunding. In other words, they spent less time on their creative work. They need to quickly learn how to become a good marketer rather than how to make their project more creative.

We found that the majority of funders were close relatives. Therefore, the idea of an anonymous crowd helping with one's campaign seems to be questionable. The average amount of money paid in the donation model was around 67.7 euro while the average amount of funders per campaign was approximately 90 euro. We found social media and emailing work best during crowdfunding process. This is not so surprising considering crowdfunding platforms rely on these forms of interaction as primary communication channels.

*Crowdfunding platform*

While there are many factors influencing the success of crowdfunding, we have a few suggestions for entrepreneurs and designers running crowdfunding platform. During the second step of our research, we aim to better inform potential founders about what the crowdfunding process involve before they start their own projects. We designed a toolkit particularly for crowdfunding founders. In our toolkits, we selected 5 common used crowdfunding models: donation model, equity model, loans model, pre-order model, and subscription model. These models are the commonly used crowdfunding models based on our interviews and surveys from crowdfunding founders. Besides, the founders also need to consider the rewards that they can offer while setting up crowdfunding projects, for example, symbolic, product, shares or none.

There are more than 100 crowdfunding platforms in the Netherlands. Some crowdfunding platforms host small budgets best (less than 5000 euro) while others best accommodate big budgets (>10000). Knowing these trends, as well as the specific budget needed, prospective founders can better decide on which platform to crowdfund. Platforms with small budgets make it easier to crowdfund from one's personal network or within a niche target (as they also need less funders), while platforms that host big budgets (that need more funders but also big funders) will more likely have a bigger database of funders or partnerships with other funding institutions. Each has its advantages, depending on what the needs of the campaign are.

If the prospective campaigners know the budget they need, they also know what financial contribution they can expect on average. For example, how many funders they need to gather. This helps them analyze the potential funders from their own network and consider how much outreach they need to make to external funders. Arguably, depending on the crowdfunding platforms and crowdfunding model, there are funders who contribute with a relatively small amount and others who contribute a far greater amount. People donate small amounts in the donation model but are inclined to invest or lend three to four times more in loan /equity models.

If we look at the role the platform plays in the relationship between founders and funders in establishing trust or helping founders choose their platform, we see little effort is made in the way they design interaction.

*Demands to the platform from the founders perspective:*

Funders can be used to further explore the development of a crowdfunding project. For example, they can help with interaction channels, making calls for participation, marketing or meetings etc.

In the interviews with some of the founders we found that they were not only interested in having their projects funded but also using the expertise of the crowd in order to improve their products. A form of crowdsourcing can take place. In other words, this means that the platform could be a channel for interaction between funders and founders. The need for multiple interactions through different channels in order to get a richer picture of the campaign could be a ground for more trust in the way funders can be involved. It was made clear that in mediated presence processes of attribution, synchronization and adaptation take place all the time (Nevejan, 2007; Steels, 2006). Because the senses have limited input and output in mediated presence — it is not the context generally the connection itself that matters — these processes of attribution, synchronization and adaptation can become very powerful (Nevejan, 2007). Therefore, on a platform different forms of attribution and synchronization should be made available to create more trust. The other form of trust that is created on the platform is witnessed presence. The connection with the funders; platforms need to center the story of the campaign in a donation model (decision influencing).

Through the video founders are claiming the need for their project that are witnessed by the possible funders. The more the founders can be contacted in different ways the more trustworthy their video will become (Nevejan, 2007). Although preliminary results of the survey indicate that funders are not that interested in connecting to other funders in a campaign this option could still be interesting to explore. Based on the interactions on social media the opinion of people you associate with can be seen as a reconfirming action. It creates possibilities to check your own decision ground and strengthen your opinion.

*Demands from the funders' point of view for a platform*

Based on what we have discovered, there are several demands from the funders' point of view for a platform. First of all, the funders need to make up their mind to fund a campaign. Interaction via Internet is asynchronical and virtual in terms of time and space. On crowdfunding platforms donation is often asked for campaigns that have an ethical element in it. Questions are often asked about to prospect funders what would be the good thing to do. For establishing the trust worthiest environment when ethical issues are at hand, a natural presence is the most effective. It is also clear that for designing a communication process like that on platforms it is necessary to see the virtual and physical interaction as a single communication process. Presence is edited and framed by technology and it is also edited and interpreted within these frameworks by people using the technology. Mediated environments that offer both information and communication facilities are therefore more attractive. The more layers of consciousness that can be addressed, the stronger the presence experience. In particular, funders need a contextualization of the campaign through different media channels on the platform when they want to make decisions.

Second, the funders would like to find out what others think of a campaign. Chatting with other funders could help them make up their mind about a campaign as so often happens nowadays in social media. People are familiar with these forms of communication, which might help them to trust a campaign. In the case of financial products being offered through the platform this form of interaction even seems compulsory based on preliminary findings in our survey.

Last but not least, they want to strengthen the bond with a campaign. The design of presence as indicated above is strongly influenced by technology. Social structures like the one in crowdfunding rely heavily on the presence of the other and yourself within this technological environment. To create trust is linked to how we design presence. The time issue of how the campaign is run is important to relate from the funders perspective. The better the updates with the latest news on how progress is being made the more chance there is to create trust among the funders participating in the campaign.

**Discussions**

In this paper, we aim at discovering the mechanisms that both founders and funders might come across during a crowdfunding process, we would like to provide practical recommendations on how to build up trust between founders and funders. In order to answer our research question, we used interviews for founders, surveys for both founders and funders. Unfortunately, due to the limitation of our research time, we couldn't finish our data collection. However, we still find a few interesting results based on our preliminary studies.

Crowdfunding is commonly hailed as the new, democratic and transparent finance model for creative industries. Crowdfunding is a business model, even when it's based on donations. It offers the same visual space and interface for all projects. Once a project has started, and even long before that, the people behind it are left with little to no time to think about the work itself or its value. While making creative workers independent of or less dependent on traditional funding sources, crowdfunding pins them to its business dynamics. It is a business model; perhaps one more complicated than traditional applications for funds or sponsorships. It comes with an extra baggage of time, free effort, multiple roles, and reliance on their own networks. But is also brings much added value, which explains why people would try it again. Our paper aims at help crowdfunding founders and funders with building trust between each other. Therefore, we would like to highlight our findings during our data collection process.

From the perspective of the founders, they are lack of crowdfunding knowledge. They are normally recommended by friends or colleagues for the particular crowdfunding platform. Even though most of the founders feel more independent when they start a crowdfunding project, most of the founders are unlikely to use crowdfunding platform next time. Most of the founders spent their time on promoting the project, which is also the most demanding part for crowdfunding. The founders need to learn how to become a good marker rather than how to make the project more creative. In addition, during the whole crowdfunding process, the founders are tending to use the knowledge of the crowd. They had to rely on their personal network during the whole crowdfunding process in order to reach the information and resources they need.

From the perspective of the funders, we found most of them prefer to have face to face meetings with the founders. However, they still want to interact with the founders through emails or crowdfunding platform. We found 85% of the funders are interested in the story behind a crowdfunding project. The personal interests and potential of the projects are two important factors for funders to select a crowdfunding project. We found the funders seldom contact the founders of the projects. Even worse, they never contact the other funders. We presume that the frequent social activities might help them improve the success rate of a crowdfunding project.

In order to build trust between founders and funders, we suggest the entrepreneur of the platforms consider the design choice, which might influence the decisions of funders during crowdfunding process. In most of the cases, unsuccessful campaigns are rarely visible on crowdfunding platforms. Visitors can only search and view what the platform's browsing filters allow them to. In other words, most of the browsing filters are designed to display only positive results and successful crowdfunding projects. These design choices—the overshadowing or complete deletion of unsuccessful projects, made it impossible for the founders and funders to get actual data for their projects.

According to our study, platform is lack of transparency in terms of crowdfunding data, business model as well as the way platforms are designed, their inner dynamics. The platforms, if equipped with good inner metrics, could provide interesting data on crowdfunding dynamics, patterns, user behavior etc. However, these data rarely reaches public surface. Platforms tend to share these insights with campaigners alone in exchange for a fee. These design choices need to be reconsidered in order to more truthfully reflect what is really happening there.

Most importantly, the platform should be built up based on the demands of founders and funders. According to the founders, the platform should be available to multiple communication channels. Presence should be considered as one of the factor in order to build trust between founders and funders. According to the funders, the funders need to make up their mind to fund a campaign. A natural presence is the most effective to them in terms of interaction as well, this is the same for the founders of the crowdfunding project. The funders are also interested in communicating with other funders, however, this is not included in most of the crowdfunding platform. In addition, the funders would like to receive timely information about an ongoing project.

In conclusion, while crowdfunding does not rely on 'traditional intermediaries', it does seem to rely on crowdfunding platforms as alternative intermediaries. The latter heavily shape how crowdfunding is developing and what cultural shifts are caused. The results we presented in our paper were based on our preliminary study. In the next step, we would like to further analyze the data we collected for this research project.

Appendix: Answer of the interview

1. Which of the following crowdfunding model do you like best?

| # | Answer | | Response | % |
|---|---|---|---|---|
| 1 | Donation model | | 4 | 31% |
| 2 | Pre-purchase model | | 6 | 46% |
| 3 | Loan mode | | 1 | 8% |
| 4 | Investment model | | 2 | 15% |
| | Total | | 13 | 100% |

2. As a funder, which of the following resources do you find more reliable about a crowdfunding project?

| # | Answer | | Response | % |
|---|---|---|---|---|
| 1 | Through social media | | 1 | 8% |
| 2 | Through website | | 4 | 31% |
| 3 | Personal meetings with initiators | | 7 | 54% |
| 4 | Email | | 0 | 0% |
| 5 | Traditional media | | 1 | 8% |
| | Total | | 13 | 100% |

**3. After being informed about a crowdfunding campaign, which of the following information can stop you from funding a project immediately?**

| # | Answer | | Response | % |
|---|---|---|---|---|
| 1 | Tittle of the project | | 3 | 25% |
| 2 | Amount of money asked | | 4 | 33% |
| 3 | Pictures of the products/services | | 2 | 17% |
| 4 | Reward of the project | | 3 | 25% |
| | Total | | 12 | 100% |

**4. Which feature on the crowdfunding platform stimulates your donation most?**

| # | Answer | | Response | % |
|---|---|---|---|---|
| 1 | The video | | 3 | 27% |
| 2 | The reward | | 3 | 27% |
| 3 | The interaction possibility with the founders/other funders | | 3 | 27% |
| 4 | The percentage of the progress for the funding project | | 2 | 18% |
| | Total | | 11 | 100% |

**5. What do you want to know most from a crowdfunding video?**

| # | Answer | | Response | % |
|---|---|---|---|---|
| 1 | Introduction of the founders | | 0 | 0% |
| 2 | Introduction of the products or services | | 2 | 15% |
| 3 | The story behind the crowdfunding project | | 11 | 85% |
| 4 | Other | | 0 | 0% |
| | Total | | 13 | 100% |

### 6. Which of the following rewards is the most important for you to fund a project ?

| # | Answer | | Response | % |
|---|---|---|---|---|
| 1 | Return of your investment (interests) | | 3 | 23% |
| 2 | The participation in the crowdfunding project | | 7 | 54% |
| 3 | If a third party is involved for the guarantee | | 1 | 8% |
| 4 | Others | | 2 | 15% |
| | Total | | 13 | 100% |

### 7. In whatway do you prefer to interact with a founder?

| # | Answer | | Response | % |
|---|---|---|---|---|
| 1 | Through emails | | 6 | 46% |
| 2 | Through social media | | 1 | 8% |
| 3 | By telephone | | 0 | 0% |
| 4 | Through crowdfunding platform | | 6 | 46% |
| 5 | Personal meeting (offline) | | 0 | 0% |
| | Total | | 13 | 100% |

### 8. On a crowdfunding campaign website, how important do you think is it to interact with other funders?

| # | Answer | | Response | % |
|---|---|---|---|---|
| 1 | Very important | | 0 | 0% |
| 2 | Important | | 3 | 23% |
| 3 | Moderately important | | 3 | 23% |
| 4 | Of little importance | | 6 | 46% |
| 5 | Unimportant | | 1 | 8% |
| | Total | | 13 | 100% |

9. **What following factor do you think is the most important for you to indicate the progress of the crowdfunding project?**

| # | Answer | Response | % |
|---|---|---|---|
| 1 | The percentage of the progress | 8 | 62% |
| 2 | The amount of participants | 2 | 15% |
| 3 | The average amount of money by the funders (already funded) | 1 | 8% |
| 4 | The time left | 2 | 15% |
|   | Total | 13 | 100% |

10. **To what extend does the presence of a founder in the video of a crowdfunding campaign make you feel comfortable to trust him/her?**

| # | Answer | Response | % |
|---|---|---|---|
| 1 | Very important | 5 | 38% |
| 2 | Important | 7 | 54% |
| 3 | Moderately important | 1 | 8% |
| 4 | Of little importance | 0 | 0% |
| 5 | Unimportant | 0 | 0% |
|   | Total | 13 | 100% |

**11. On a crowdfunding campaign website, how important do you think is it to get timely information from the founder of a project?**

| # | Answer | Response | % |
|---|---|---|---|
| 1 | Very important | 0 | 0% |
| 2 | Important | 10 | 77% |
| 3 | Moderately important | 3 | 23% |
| 4 | Of little importance | 0 | 0% |
| 5 | Unimportant | 0 | 0% |
|   | Total | 13 | 100% |

**12. In a crowdfunding campaign, how important do you think is it to have shared interests with a founder of a project?**

| # | Answer | Response | % |
|---|---|---|---|
| 1 | Very important | 2 | 15% |
| 2 | Important | 4 | 31% |
| 3 | Moderately important | 4 | 31% |
| 4 | Of little importance | 2 | 15% |
| 5 | Unimportant | 1 | 8% |
|   | Total | 13 | 100% |

**13. What are the most important factors for you to select a crowdfunding project?**

| # | Answer | | Response | % |
|---|---|---|---|---|
| 1 | Personal Interests | | 6 | 46% |
| 2 | The potential of the projects | | 4 | 31% |
| 3 | Trust about the founder of the projects | | 2 | 15% |
| 4 | Recommendation by friends | | 0 | 0% |
| 5 | Recommendation by influencers | | 0 | 0% |
| 6 | Others | | 1 | 8% |
| | Total | | 13 | 100% |

**14. How many times have you contacted the founder of the project in your past funding experience on average?**

| # | Answer | | Response | % |
|---|---|---|---|---|
| 1 | Never | | 5 | 38% |
| 2 | 1-5 times | | 8 | 62% |
| 3 | 6-20 times | | 0 | 0% |
| 4 | Above 20 times | | 0 | 0% |
| | Total | | 13 | 100% |

**15. How many times have you contacted other funders of the project in your past funding experience on average?**

| # | Answer | | Response | % |
|---|---|---|---|---|
| 1 | Never | | 13 | 100% |
| 2 | 1-5 times | | 0 | 0% |
| 3 | 6-20 times | | 0 | 0% |
| 4 | Above 20 times | | 0 | 0% |
| | Total | | 13 | 100% |

**16. How many times have you contacted the platform for any questions or problems on average?**

| # | Answer | Response | % |
|---|---|---|---|
| 1 | Never | 6 | 50% |
| 2 | 1-5 times | 6 | 50% |
| 3 | 6-20 times | 0 | 0% |
| 4 | Above 20 times | 0 | 0% |
|   | Total | 12 | 100% |

**17. How important is your relationship with the founder in a crowdfunding campaign?**

| # | Answer | Response | % |
|---|---|---|---|
| 1 | Very important | 2 | 15% |
| 2 | Important | 3 | 23% |
| 3 | Moderately important | 3 | 23% |
| 4 | Of little importance | 5 | 38% |
| 5 | Unimportant | 0 | 0% |
|   | Total | 13 | 100% |